\documentclass[aps,twocolumn,pra,superscriptaddress]{revtex4}
\usepackage{epsfig,graphicx,times}
\usepackage{amstext}
\usepackage{amsmath}
\usepackage{lipsum}
\usepackage{amssymb}
\usepackage{float}
\usepackage{graphicx}
\usepackage{latexsym}
\usepackage{bm}
\usepackage{appendix}
\usepackage[colorlinks,citecolor=blue, linkcolor=blue,hyperindex,CJKbookmarks]{hyperref}

\begin{document}

\title{Agnostic Parameter Estimation with Large Spins}
\author{Huining Zhang}
\affiliation{Center for Quantum Sciences and School of Physics, Northeast Normal University, Changchun 130024, China}
\author{X. X. Yi\footnote{yixx@nenu.edu.cn}}
\affiliation{Center for Quantum Sciences and School of Physics, Northeast Normal University, Changchun 130024, China}
\affiliation{Center for Advanced Optoelectronic Functional Materials Research, and Key Laboratory for UV Light-Emitting Materials and Technology of Ministry of Education, Northeast Normal University, Changchun 130024, China}

\date{\today}

\begin{abstract}
The quantum Fisher information of a quantum state  with respect to  a certain parameter quantifies the sensitivity of the quantum state to changes in that parameter. Maximizing the quantum Fisher information is essential for achieving the optimal estimation precision of quantum sensors. A typical quantum sensor involves a qubit(e.g. a spin-$\frac 1 2$) probe undergoing an unknown rotation, here the unknown rotation angle is the parameter to be estimated. A  well known limitation is that if the rotation axis is unknown, the maximal quantum Fisher information is impossible to attain. This limitation has been lifted recently  by leveraging entanglement between the probe qubit and an ancilla qubit. Namely, through measurement of the ancilla after the axis is revealed, one can prepare the probe that is optimal for any unknown rotation axis. This proposal, however, works only for a spin-$\frac 1 2$. Considering large spin probes can achieve a larger quantum Fisher information, offering enhanced metrological advantage, we here  utilize the entanglement between a large spin probe and an ancilla to achieve optimal quantum Fisher information for estimating the rotation angle, without prior knowledge of the rotation axis. Different from the previous spin-$\frac 1 2$ case, achieving the optimal precision with large spins generally requires post-selection, resulting in a success probability dependent on the dimension of the Hilbert space. Furthermore, we extend the encoding    state from the maximally entangled case to general entangled states, showing that optimal metrology can still be achieved with a certain success probability.
\end{abstract}
\maketitle

\section{INTRODUCTION}
Based on the theory of quantum parameter estimation, quantum metrology  \cite{Gio2006} has emerged as one of the most developed areas in quantum information processing in the last decades. Accurate estimation of physical parameters is crucial across numerous scientific fields, driving the rapid development of the parameter estimation theory \cite{Gio2011,Degen2017,Liu2020,Ali2014,Hou2020,Hum2013,Dorn2009,Yuan2015,Chen2024,Lu2021, Ding2021,Jing2015,Bai2023,Peng2024,Yu2023,Dem2014,Li2024,Li2023,Shao2023,Pang2014, Beau2017,Arvi2023,Song2024,Song2025}. Among them, the phase estimation theory\cite{Hel1976,Hol1982} plays an important role, since it solves the problem of precisely estimating the eigenphase of a unitary operator, a task that is exponentially hard for classical computer in many cases.  The phase estimation theory  determines unknown physical parameters encoded in quantum systems through a unitary transformation, typically $U=e^{-i\beta H}$ \cite{Gio2006,Dorn2009,Song2024}. Here, $H$ is the Hermitian generator, and the parameter $\beta$ (such as the strength of an unknown field) represents the phase to be estimated. By applying $U$ to a probe encoding  state and performing a suitable measurement, one can obtain an estimate for $\beta$.

The classical Fisher information(FI) quantifies the probability  sensitivity to small changes in the estimated parameter. A larger Fisher information implies that one can estimate the parameter with a higher precision. Maximizing the Fisher information over all possible quantum measurements defines the quantum Fisher information (QFI) \cite{Hel1976,Hol1982,Hub1992,Hub1993,Brau1994}, which characterizes the sensitivity of a quantum state to the parameter perturbations and provides a lower limit for precision, known as the quantum Cram$\acute{\text e}$r-Rao bound \cite{Brau1994,Cramer1946,Brau1996}.
The optimal probe encoding state that maximizes the QFI is an equally weighted superposition of the eigenvectors corresponding to the maximum and minimum eigenvalues of $H$ \cite{Gio2006,Gio2011,Brau1994,Brau1996}. Therefore, a fundamental requirement of standard quantum metrology protocols is complete prior knowledge of the generator $H$, as this knowledge determines the structure of the optimal encoding state. If $H$ is not known beforehand, the conventional phase estimation method fails.

Recently, the authors of \cite{Arvi2023,Song2024} introduced a phase-estimation protocol designed for scenarios where information about the generator $H$ is available only after  the parameter-encoding   unitary has acted.  Their protocol entangles a probe with an ancilla in the maximum entangled state. Once $H$ is known, a measurement on the ancilla effectively prepares (or ``updates'') the probe into an optimal state, leveraging a mathematical equivalence to closed timelike curves \cite{Godel1949,Mor1988,Svet2011,Lloyd2011,Lloy2011}. This establishes a powerful framework for designing metrological protocols that use entanglement to circumvent the need for prior information. The advantage of this approach was demonstrated experimentally with superconducting qubits \cite{Song2024}. We should notice that  both these studies focused on qubits (spin-$\frac{1}{2}$). For a spin probe under linear coupling with fields(e.g., $H\propto\bm{S}\cdot\bm{n}$), the maximum quantum Fisher information scales quadratically with the spin quantum number $s$. Therefore, employing large spins ($s>\frac{1}{2}$) offers a useful route to enhance the sensitivity.

In this work, we introduce a theoretical framework which generalize the protocol with spin-$\frac 1 2$ to that with large spins. We address how probe-ancilla entanglement can be harnessed to achieve the maximum QFI for estimating $\beta$ when $H$ is completely unknown a priori, and we elucidate the key differences between  ours and the  spin-$\frac 1 2$ case.
We find that for large spins, manipulation of probe-ancilla entanglement enables optimal estimation of the parameter $\beta$ via postselection \cite{Aha2002}, even in the complete absence of prior knowledge about the generator $H$. A key difference between this proposal and  the spin-$\frac 1 2$  case \cite{Arvi2023,Song2024} is that the success here is probabilistic. For an encoding  maximally entangled state, we derive the success probability that can achieve the optimal estimation and analyze its dependence on the spin dimension $m$. The condition that the success probability reaches its maximum is analyzed. Furthermore, we generalize our protocol to arbitrary (non-maximal) entangled encoding  states, and show that optimal $H$-agnostic estimation remains valid, with a success probability now depending on both the specific form of the entangled state and the generator $H$.

This work is organized as follows. In Sec.~\ref{sec1}, we briefly review the foundational concepts of quantum metrology relevant to our studies. Sec.~\ref{sec2}  investigates how probe-ancilla entanglement enables optimal parameter estimation with large spins, regardless of the unknown rotation axis. Here the ancilla states in the probe-ancilla entangled state are mutually orthogonal. Specific examples utilizing this protocol are also detailed in this section. In Sec.~\ref{sec3}, we gives the entanglement sensing protocol in large spins when the states of the ancilla are not mutually orthogonal. As an example, we analyze the case of spin-1 in detail. Finally, we summarize our results in Sec.~\ref{sec4}. Appendix \ref{appendixA} is provided as supplemental materials to Sec.~\ref{sec3}.

\section{BASIC THEORY OF QUANTUM METROLOGY}\label{sec1}

We begin by reviewing key concepts of quantum parameter estimation theory that we will use in later discussions. Consider a common scenario where an unbiased estimator $\hat{\beta}$ is obtained from $N$ measurement outcomes. The Fisher information (FI), denoted $F$, quantifies the sensitivity of the outcome probabilities to small changes in $\beta$. $F$ limits the variance of estimator via the Cram$\acute{\text{e}}$r-Rao bound \cite{Brau1994,Cramer1946,Brau1996},
\begin{equation}
\text{var}(\hat{\beta})\geq\frac{1}{NF}\geq\frac{1}{N\mathcal{F}}.
\label{eq1}
\end{equation}
The FI is upper bounded by the quantum Fisher information (QFI),  which is defined as the maximum FI attainable over all possible quantum measurements \cite{Brau1994}.

In our manuscript, we consider a parameter estimation scenario where the parameter $\beta$ is encoded via a unitary $U=e^{-i\beta H}$ onto an encoding    pure probe state $|\psi\rangle$. The estimated parameter $\beta$ is an overall factor of Hamiltonian $H$ and $H$ is independent of $\beta$. The evolved state is $|\psi_\beta\rangle = U|\psi\rangle$. If the probe is measured in a set of bases \{$|i\rangle$\}, the probability for outcome $i$ is $P_i(\beta) = |\langle i | \psi_\beta \rangle|^2$. For this measurement, the FI with respect to $\beta$ is given by
\begin{equation}
F=\sum_i\frac{(\partial_\beta P_i)^2}{P_i}.
\label{eq2}
\end{equation}
For the parametrized density matrix $\rho=|\psi_\beta\rangle\langle\psi_\beta|$, the quantum Fisher information is defined as
\begin{equation}
\mathcal{F}=\text{Tr}(\rho L^2),
\label{eq3}
\end{equation}
where $L$ is the symmetric logarithmic derivative operator, which is defined implicitly by $\partial_\beta\rho=(L\rho+\rho L)/2$ \cite{Brau1994,Brau1996}. For the evolved state $e^{-i\beta H}|\psi\rangle$, the QFI is proportional to the variance of the Hamiltonian $H$ in $|\psi\rangle$
\begin{equation}
\begin{aligned}
\mathcal{F}=4\langle\psi|\Delta H^2|\psi\rangle.
\label{eq4}
\end{aligned}
\end{equation}
To maximize the variance of $H$ (and hence the QFI), the optimal encoding  probe state is an equally weighted superposition of the eigenvectors  $|\lambda_M\rangle$ and $|\lambda_m\rangle$, which correspond to the maximum and minimum eigenvalues $\lambda_M$ and $\lambda_m$ of the Hamiltonian $H$, respectively \cite{Brau1994,Gio2011,Brau1996,Gio2006}. That is
\begin{equation}
\begin{aligned}
|\psi\rangle=\frac{1}{\sqrt{2}}(|\lambda_M\rangle+e^{i\alpha}|\lambda_m\rangle),
\label{eq5}
\end{aligned}
\end{equation}
where $\alpha$ is an arbitrary relative phase. Here $|\psi\rangle$ is the optimal encoding  state, and the maximal QFI (MQFI) is
\begin{equation}
\begin{aligned}
\mathcal{F}_{\text {max}}=(\lambda_M-\lambda_m)^2.
\label{eq6}
\end{aligned}
\end{equation}

In this work, we estimate the strength $\beta$ encoded in a single-spin probe subjected to an unknown unitary operation $U_{\bm{n}}(\beta)=e^{-i\beta\bm{S}\cdot\bm{n}}$. Here, $\bm{S} = (S_x, S_y, S_z)$ are the spin-$s$ operators acting on a Hilbert space of dimension $\mathcal H=2s+1=m$, satisfying the commutation relation $[S_i, S_j] = i\hbar\epsilon_{ijk}S_k$. The unit vector $\bm{n} = (\sin\theta, 0, \cos\theta)$ parameterizes the unknown axis of rotation. Here we mainly focus on large spins (with $\mathcal H=2s+1\geq3$). The Hamiltonian reads,
\begin{equation}
\begin{aligned}
H=\boldsymbol{S}\cdot\boldsymbol{n}=S_x\sin\theta+S_z\cos\theta.
\label{eq7}
\end{aligned}
\end{equation}
The eigenvalues of $H$ are
\begin{equation}
\begin{aligned}
E_i=s-i+1, (i=1,2,\cdots,m)
\label{eq8}
\end{aligned}
\end{equation}
and the corresponding eigenstates of \{$E_i$\} are \{$|E_i\rangle$\}. The optimal probe states can be expressed as
\begin{equation}
\begin{aligned}
|n_\pm\rangle=\frac{1}{\sqrt{2}}(|E_1\rangle\pm|E_m\rangle),
\label{eq9}
\end{aligned}
\end{equation}
which maximize the quantum Fisher information
\begin{equation}
\begin{aligned}
\mathcal{F}_{\text {max}}=(E_1-E_m)^2=4s^2.
\label{eq10}
\end{aligned}
\end{equation}
From Eq.~(\ref{eq10}) we can find that with large spins (i.e., high spin quantum number $s$) as quantum probes, the MQFI increases  significantly leading to improvements in the parameter estimation precision. This enhancement arises from the quadratic dependence of MQFI on $s$, which is superior over the achievable limit with spin-1/2 probes. Motivated by this advantage, we choose large spins as probes for quantum sensors.

It shows that conventional phase estimation requires prior knowledge about  the unitary. In the limit of a large number of trials ($N\rightarrow\infty$), all bounds in Eq.~(\ref{eq1}) can be saturated if $H$ is known. Without information about $H$, one cannot prepare the optimal encoding state in advance. This obstacle, however, can be circumvented through probe-ancilla qubit entanglement manipulation \cite{Arvi2023,Song2024}. By measuring the ancilla entangled with the probe after $U$ has acted, the probe can be projected into an optimal state for estimation, even when $H$ was initially unknown.

\section{Entanglement-Assisted Metrology with Large spins in Orthogonal Ancilla States}\label{sec2}

In this section, we extend  the protocol \cite{Arvi2023,Song2024} from qubits to large spins for ancilla in a superposition of orthogonal states.
We start with  initializing  a probe spin $A$ and an ancilla spin $B$ in an entangled   state
\begin{equation}
\begin{aligned}
|\Psi\rangle_{AB}=\sum_{i,j=1}^{m}\chi_{ij}|i-1\rangle_A\otimes|j-1\rangle_B,
\label{eq11}
\end{aligned}
\end{equation}
where \{$\chi_{ij}$\} is a set of coefficients. The orthonormal basis states $\{|i-1\rangle\}_{i=1}^m$ span a $m$-dimensional Hilbert space, where $|0\rangle=[1,0,0,\ldots]^T, |1\rangle=[0,1,0,\ldots]^T$.
Such an entangled   state admits a Schmidt decomposition \cite{Sch1907,Pere1995}
\begin{equation}
\begin{aligned}
|\Psi\rangle_{AB}=\sum_{k=1}^{m}\xi_k|u_k\rangle_A\otimes|v_k\rangle_B,
\label{eq12}
\end{aligned}
\end{equation}
where $r$ is the Schmidt rank ($1<r\leq m$), indicating the number of non-zero Schmidt coefficients $\xi_k$. $\xi_k\geq0$ and $\sum_{k}\xi_k^2=1$.  \{$|u_k\rangle_A$\} and \{$|v_k\rangle_B$\} are respectively a set of orthogonal basis vectors for $A$ and $B$. Different entangled states are characterized by different sets of Schmidt coefficients and bases, and lead to different results as we will show in the next section.

Eq.~(\ref{eq9}) shows that the form of the optimal encoding states of the probe are $|n_\pm\rangle$. They form a set of orthogonal complete bases \{$|\phi_i\rangle$\} together with the other eigenstates of the Hamiltonian. $|\phi_i\rangle$ denoted by
\begin{equation}
\begin{aligned}
|\phi_i\rangle &=
\begin{cases}
|n_+\rangle, & i=1, \\
|E_i\rangle, & i=2, \cdots, m-1, \\
|n_-\rangle, & i=m.
\end{cases}
\end{aligned}
\label{eq13}
\end{equation}
The basis vectors \{$|u_k\rangle$\} of the probe $A$ can be expanded in the bases \{$|\phi_i\rangle$\} satisfying
\begin{equation}
\begin{pmatrix}
|u_1\rangle_A \\
|u_2\rangle_A \\
\vdots \\
|u_m\rangle_A
\end{pmatrix}
=S\cdot
\begin{pmatrix}
|\phi_1\rangle_A \\
|\phi_2\rangle_A \\
\vdots \\
|\phi_m\rangle_A
\label{eq14}
\end{pmatrix}.
\end{equation}
$S=\{{S_{ki}}\}$ is a $m\times m$ transformation matrix. Substituting Eq.~(\ref{eq14}) into Eq.~(\ref{eq12}) yields
\begin{equation}
\begin{aligned}
|\Psi\rangle_{AB}&=\sum_{i=1}^{m}|\phi_i\rangle_A\otimes|\tilde{\psi}_i\rangle_B=\sum_{i=1}^{m}c_i|\phi_i\rangle_A\otimes|
\psi_i\rangle_B,\\
|\tilde{\psi}_i\rangle_B&=\sum_{k=1}^{m}\xi_k S_{ki}|v_k\rangle_B,
\label{eq15}
\end{aligned}
\end{equation}
where $c_i=\sqrt{_B\langle\tilde{\psi}_i|\tilde{\psi}_i\rangle_B}\geq0$. When $c_i\neq0$, $|\psi_i\rangle_B=|\tilde{\psi}_i\rangle_B/c_i$. Otherwise, $c_i=0$ and $|\tilde{\psi}_i\rangle$ corresponds to a zero vector in the Hilbert space.

We suppose  $H$ is initially unknown and consider that the encoding  entangled state satisfies the condition $_B\langle\psi_i|\psi_j\rangle_B=\delta_{ij}$ (for any $|\psi_i\rangle_B$ with non-zero $c_i$). Once $\bm{n}$ is revealed, a measurement of $B$ in these bases allows for the optimal encoding   state of probe $A$ to be prepared retroactively. Namely,  after the unknown  encoding unitary has acted, $\bm{n}$ is revealed.

\begin{figure}[t]
	\centering
	\includegraphics[width=0.45\textwidth]{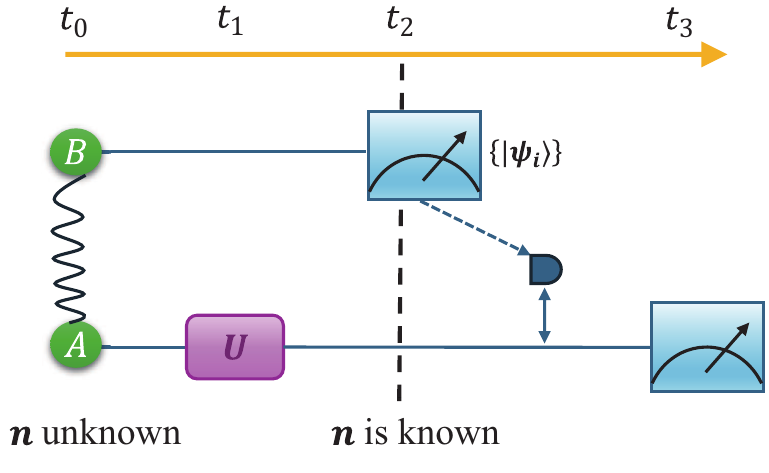}
	\caption {Circuit diagram for optimal parameter estimation without prior knowledge of the rotation axis. At $t_0$, the probe-ancilla pair are in an entangled state, while the axis $\bm n$ is unknown. The unitary $U_{\bm{n}}(\beta) = e^{-i\beta H}$ acts on the probe at $t_1$. At $t_2$, $\bm n$ is revealed. A measurement on $B$ effectively teleports to the past an optimal encoding  state for the probe $A$ via postselection.}
	\label{aqbpic0}
\end{figure}

The protocol proceeds as follows (see Fig.~\ref{aqbpic0}):

$t_0$:$\star$ $A$ and $B$ are entangled:
\begin{equation}
\begin{aligned}
|\Psi\rangle_{AB}=\sum_{k=1}^{m}\xi_k|u_k\rangle_A\otimes|v_k\rangle_B.\nonumber
\end{aligned}
\end{equation}

$t_1$:$\star$ $A$ is subject to the unknown unitary $U_{\bm{n}}(\beta)=e^{-i\beta\bm{S}\cdot\bm{n}}$.

$\phantom{t_1{:}}\star$ The joint state becomes
\begin{equation}
\begin{aligned}
U_{\bm{n}}(\beta)\otimes\hat{I}_B|\Psi\rangle_{AB}=\sum_{i=1}^{m}c_ie^{-i\beta\bm{S}\cdot\bm{n}}|\phi_i\rangle_A\otimes|
\psi_i\rangle_B.\nonumber
\end{aligned}
\end{equation}

$t_2$:$\star$ The form of $\bm{n}$ is now known.

$\phantom{t_2{:}}\star$ $B$ is measured in a set of orthogonal  bases that contain the states \{$|\psi_i\rangle_B$\}. If the outcome is $|\psi_1\rangle$ or $|\psi_m\rangle$, the blocking element in Fig.~\ref{aqbpic0} is removed. If not, the blocking element will destroy $A$.

$\phantom{t_2{:}}\star$ Depending on the outcome, the state of $A$ becomes $e^{-i\beta\bm{S}\cdot\bm{n}}|n_\pm\rangle$ with a certain probability. The probability  is given by
\begin{equation}
\begin{aligned}
p=c_1^2+c_m^2=\sum_{k=1}^{m}\xi_k^2(|S_{k1}|^2+|S_{km}|^2).
\label{eq16}
\end{aligned}
\end{equation}

$t_3$:$\star$ $A$ is measured in the optimal basis $\{|\phi_i\rangle\}$.

This scheme works for  phase estimation with large spins  and relaxes the requirement for prior knowledge. In the following, we proceed to examine a series of representative cases as examples.

\subsection{The maximally entangled encoding   state}\label{sec2a}
In this example,  we consider the case where the probe $A$ and ancilla $B$ are prepared in a maximally entangled encoding    state
\begin{equation}
\begin{aligned}
|\Psi\rangle_{AB}=\frac{1}{\sqrt{m}}\sum_{k=1}^{m}|u_k\rangle_A\otimes|v_k\rangle_B,
\label{eq17}
\end{aligned}
\end{equation}
where the dimension of the Hilbert space is $\mathcal{H}_A=\mathcal{H}_B=m$. \{$|u_k\rangle_A$\} and \{$|v_k\rangle_B$\} are respectively orthogonal complete bases for $A$ and $B$, and their forms are independent of $\bm{n}$. According to Eq.~(\ref{eq14}), $|u_k\rangle_A$ can be expressed as linear combinations of the bases \{$|\phi_i\rangle$\}
\begin{equation}
\begin{aligned}
|u_k\rangle_A=\sum_{i=1}^{m}U_{ki}|\phi_i\rangle_A.
\label{eq18}
\end{aligned}
\end{equation}
It is easy to prove that $U=\{U_{ki}\}$ is a unitary matrix and satisfies $UU^\dagger=U^\dagger U=I$. Eq.~(\ref{eq17}) can be rewritten as
\begin{equation}
\begin{aligned}
|\Psi\rangle_{AB}=\frac{1}{\sqrt{m}}\sum_{i=1}^{m}|\phi_i\rangle_A\otimes|\psi_i\rangle_B,
\end{aligned}
\end{equation}
where
\begin{equation}
\begin{aligned}
|\psi_i\rangle_B=\sum_{k=1}^{m}U_{ki}|v_k\rangle_B.
\label{eq20}
\end{aligned}
\end{equation}
The unitarity of $U$ ensures that the states ${|\psi_i\rangle_B}$ are orthogonally normalised: $_B\langle\psi_i|\psi_j\rangle_B=\sum_{kl}U_{ki}^{*}U_{lj}\langle v_k|v_l\rangle=\sum_{k}U_{ki}^{*}U_{kj}=\delta_{ij}$. Therefore, a measurement of $B$ in these bases then projects $A$ onto the optimal state $e^{-i\beta\bm{S}\cdot\bm{n}}|n_\pm\rangle$ with the success probability $p$ given by Eq.~(\ref{eq16})
\begin{equation}
\begin{aligned}
p=c_1^2+c_m^2=\frac{2}{m}.
\end{aligned}
\end{equation}
Here we work in the basis $|u_k\rangle=|v_k\rangle=|k-1\rangle$. The Hamiltonian $H$ in Eq.~(\ref{eq7}) is a real symmetric matrix, hence, the matrix of its eigenvectors ${|E_i\rangle}$ can be taken  real. Similarly, the matrix of basis ${|\phi_i\rangle}$ (which is related to ${|E_i\rangle}$) is also real. This together with Eq.~(\ref{eq18}) implies that $U^\dagger = U^T$, and
\begin{equation}
\begin{aligned}
(U^\dagger U)_{ij}=\sum_kU_{ik}^\dagger U_{kj}=\sum_kU_{ki}U_{kj}=\delta_{ij}.
\end{aligned}
\end{equation}
Then it can be found that
\begin{equation}
\begin{aligned}
|\psi_i\rangle_B&=\sum_{k=1}^{m}U_{ki}|k-1\rangle_B=\sum_{k=1}^{m}U_{ki}(\sum_{j=1}^{m}U_{kj}|\phi_j\rangle_B)\\
&=\sum_{j=1}^{m}(\sum_{k=1}^{m}U_{ki}U_{kj})|\phi_j\rangle_B=|\phi_i\rangle_B.
\end{aligned}
\end{equation}

For the maximally entangled state, the structure of $|\Psi\rangle_{AB}$ does not depend on the basis \{$|u_k\rangle_A$\} and \{$|v_k\rangle_B$\}. The proposal proceeds as follows.

$t_0$:$\star$ $A$ and $B$ are entangled:
\begin{equation}
\begin{aligned}
|\Psi\rangle_{AB}=\frac{1}{\sqrt{m}}\sum_{k=1}^{m}|k-1\rangle_A|k-1\rangle_B=
\frac{1}{\sqrt{m}}\sum_{i=1}^{m}|\phi_i\rangle_A|\phi_i\rangle_B.\nonumber
\end{aligned}
\end{equation}

$t_1$:$\star$ $A$ is subject to the unitary $U_{\bm{n}}(\beta)=e^{-i\beta\bm{S}\cdot\bm{n}}$.

$\phantom{t_1{:}}\star$ The joint state becomes $\frac{1}{\sqrt{m}}\sum_{i=1}^{m}e^{-i\beta\bm{S}\cdot\bm{n}}|\phi_i\rangle_A|\phi_i\rangle_B$

$t_2$:$\star$ The form of $\bm{n}$ is now revealed.

$\phantom{t_2{:}}\star$ $B$ is measured in the orthogonal complete bases \{$|n_\pm\rangle, |E_i\rangle$ $\scriptstyle (i=2, \cdots, m-1)$\}. If the outcome is $|n_\pm\rangle$, the blocker is removed.

$\phantom{t_2{:}}\star$ $A$ collapse to the metrologically optimal state $e^{-i\beta\bm{S}\cdot\bm{n}}|n_\pm\rangle$.

$t_3$:$\star$ $A$ is measured in the optimal $\{|\phi_i\rangle\}$ basis.
\begin{figure}[t]
	\centering
	\includegraphics[width=0.5\textwidth]{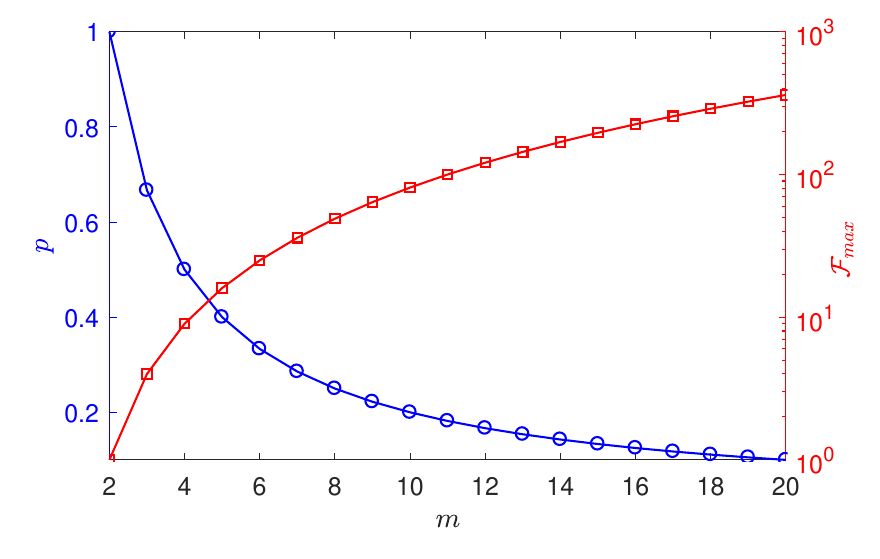}
	\caption {The blue curve represents the probability of successfully collapsing to the metrologically optimal state for the probe $A$ as a function of the Hilbert space dimension $m$. The red curve represents the maximum quantum Fisher information under the optimal state as a function of $m$.}
	\label{aqbpic1}
\end{figure}

For the large spin initialized in a maximally entangled state, a measurement (postselection) on the ancilla in the basis spanned by ${|n_\pm\rangle}$ and the other eigenstates ${|E_i\rangle}$ of $H$ can project the probe onto an optimal state for estimating $\beta$. The postselection success probability  depends on the dimension $m$. As shown in Fig.~\ref{aqbpic1}, this probability decreases with increasing spin quantum number $s$.  Meanwhile, the MQFI $\mathcal{F}_{\text {max}}$ attained with this optimal probe state increases with the dimension of the spin probe. This implies that  a higher achievable precision for estimating the parameter $\beta$ can be probabistically  obtained.

We interpret these results using the language of closed timelike curves (CTCs) \cite{Arvi2023,Song2024}. At $t_0$, a probe spin $A$ and an ancilla spin $B$ are prepared in the maximally entangled state. The probe is expecting the optimal encoding state  to be received from the future(i.e., after the measurement of the ancilla). Then the probe undergoes a unitary rotation $U_{\bm{n}}(\beta)$ about an unknown axis $\bm{n}$. After the interaction, $\bm{n}$ is revealed \cite{Arvi2023,Song2024}. A measurement projects the ancilla onto $|n_\pm\rangle$ with success probability $p$. This post-selected state is effectively utilized as if it were sent back to time $t_0$ to serve as the probe's encoding  state. Thus, the probe is prepared in the optimal state retroactively, rotated by $U_{\bm{n}}(\beta)$, and finally, at $t_3$, subjected to an optimal measurement in the basis $\{|\phi_i\rangle\}$.

\subsection{The condition for the maximal success probability}\label{sec2b}
The maximum success probability $p=1$ is achieved, according to Eq.~(\ref{eq16}), when $c_1^2+c_m^2=1$. We now examine the conditions under which this occurs. $p=1$ implies that the coefficients in Eq.~(\ref{eq15})
\begin{equation}
\begin{aligned}
c_2^2=\cdots=c_{m-1}^2=0,
\label{eq24}
\end{aligned}
\end{equation}
and satisfy $\langle\psi_1|\psi_m\rangle=0$ in our scheme. The entanglement state becomes from Eq.~(\ref{eq15})
\begin{equation}
\begin{aligned}
|\Psi\rangle_{AB}&=c_1|n_+\rangle_A|{\psi}_1\rangle_B+c_m|n_-\rangle_A|{\psi}_m\rangle_B\\
&=\xi_1|u_1\rangle_A|v_1\rangle_B+\xi_2|u_2\rangle_A|v_2\rangle_B.
\label{eq25}
\end{aligned}
\end{equation}
This corresponds to the following specific form in the Schmidt decomposition of Eq.~(\ref{eq12}): The probe bases are $|u_1\rangle = |n_+\rangle$ and $|u_2\rangle = |n_-\rangle$; the ancilla bases are $|v_1\rangle = |\psi_1\rangle$ and $|v_2\rangle = |\psi_m\rangle$; and the coefficients are $c_1 = \xi_1$, $c_m = \xi_2$, with $\xi_1, \xi_2 > 0$ and $\xi_1^2 + \xi_2^2 = 1$.

To be specific,  $|v_1\rangle=|n_-\rangle$, $|v_m\rangle=-|n_+\rangle$, and they follow the relation with the orthogonal basis
\begin{equation}
\begin{pmatrix}
|n_+\rangle \\
|n_-\rangle \\
\end{pmatrix}
=\mathcal M
\begin{pmatrix}
|0\rangle \\
|1\rangle \\
\vdots \\
|m-1\rangle
\label{eq26}
\end{pmatrix}.
\end{equation}
$\mathcal M$ is a $2\times m$ transformation matrix. $|n_+\rangle=\sum_{i=1}^{m}{\mathcal M}_{1i}|i-1\rangle$ and $|n_-\rangle=\sum_{i=1}^{m}{\mathcal M}_{2i}|i-1\rangle$. Plugging Eq.~(\ref{eq26}) into Eq.~(\ref{eq25}), we get the coefficients in Eq.~(\ref{eq11})
\begin{equation}
\begin{aligned}
\chi_{ij}=\xi_1{\mathcal M}_{1i}{\mathcal M}_{2j}-\xi_2{\mathcal M}_{2i}{\mathcal M}_{1j}.
\label{eq27}
\end{aligned}
\end{equation}
Initially, the probe and ancilla are prepared in an entangled state. A unitary $U_{\bm{n}}(\beta)$, encoding   the unknown parameter $\beta$ about an unknown axis $\bm{n}$, acts on the probe. The coefficients $\chi_{ij}$ are adjustable. If, despite $\bm{n}$ being unknown, the chosen $\chi_{ij}$ equal the specific values given in Eq.~(\ref{eq27}), then the success probability reaches its maximum of 1. This represents an ideal scenario where the adjustable parameters coincidentally match those required for the  unknown $\bm{n}$.

The protocol proceeds as follows.

$t_0$:$\star$ $A$ and $B$ are entangled:
\begin{equation}
\begin{aligned}
|\Psi\rangle_{AB}=\sum_{i,j=1}^{m}\chi_{ij}|i-1\rangle_A\otimes|j-1\rangle_B.\nonumber
\end{aligned}
\end{equation}

$t_1$:$\star$ $A$ is subject to the unitary $U_{\bm{n}}(\beta)=e^{-i\beta\bm{S}\cdot\bm{n}}$.

$\phantom{t_1{:}}\star$ The joint state becomes
\begin{equation}
\begin{aligned}
\xi_1e^{-i\beta\bm{S}\cdot\bm{n}}|n_+\rangle_A|n_-\rangle_B-\xi_2e^{-i\beta\bm{S}\cdot\bm{n}}|n_-\rangle_A|n_+\rangle_B.
\nonumber
\end{aligned}
\end{equation}

$t_2$:$\star$ The form of $\bm{n}$ is now known.

$\phantom{t_2{:}}\star$ $B$ is measured in the bases \{$|n_\pm\rangle, |E_i\rangle$ $\scriptstyle (i=2, \cdots, m-1)$\}.

$\phantom{t_2{:}}\star$ $A$ collapse to $e^{-i\beta\bm{S}\cdot\bm{n}}|n_\pm\rangle$ with the probability $p=1$.

$t_3$:$\star$ $A$ is measured in the optimal $\{|\phi_i\rangle\}$ basis.

 Our general framework for large spins naturally recovers the earlier qubit results given in \cite{Arvi2023,Song2024}. For a spin-$\frac{1}{2}$ probe, the unitary is $U = e^{-i\beta (\bm{\sigma}\cdot\bm{n})/2}$. In terms of  the unknown generator, it reads
\begin{equation}
\begin{aligned}
H=\frac{1}{2}\boldsymbol{\sigma}\cdot\boldsymbol{n}=\frac{1}{2}(\sigma_x\text{sin}\theta+\sigma_z\text{cos}\theta),
\label{eq28}
\end{aligned}
\end{equation}
where $\sigma_x$ and $\sigma_z$ are Pauli operators. The eigenvalues of $H$ are $\pm\frac{1}{2}$. The corresponding eigenstates are
\begin{equation}
|+\frac{1}{2}\rangle
=
\begin{pmatrix}
\text{cos}\frac{\theta}{2} \\
\\
\text{sin}\frac{\theta}{2}
\end{pmatrix},
|-\frac{1}{2}\rangle
=
\begin{pmatrix}
-\text{sin}\frac{\theta}{2} \\
\\
\text{cos}\frac{\theta}{2}
\label{eq29}
\end{pmatrix}.
\end{equation}
The optimal states $|n_\pm\rangle$ in Eq.~(\ref{eq29}) can be rewritten in the form of Eq.~(\ref{eq26})
as
\begin{equation}
\begin{pmatrix}
|n_+\rangle \\
\\
|n_-\rangle
\end{pmatrix}
=\frac{1}{\sqrt{2}}
\begin{pmatrix}
a-b & a+b\\
\\
a+b & b-a
\end{pmatrix}
\begin{pmatrix}
|0\rangle \\
\\
|1\rangle
\label{eq30}
\end{pmatrix},
\end{equation}
where $a=\text{cos}\frac{\theta}{2}$ and $b=\text{sin}\frac{\theta}{2}$.
From Eq.~(\ref{eq27}), we get the coefficients in Eq.~(\ref{eq11})
\begin{equation}
\begin{aligned}
\chi_{11}&=\frac{1}{2}(\xi_1-\xi_2)\cos\theta,\\
\chi_{12}&=-\frac{\xi_1}{2}(1-\sin\theta)-\frac{\xi_2}{2}(1+\sin\theta),\\
\chi_{21}&=\frac{\xi_1}{2}(1+\sin\theta)+\frac{\xi_2}{2}(1-\sin\theta),\\
\chi_{22}&=-\frac{1}{2}(\xi_1-\xi_2)\cos\theta.
\label{eq31}
\end{aligned}
\end{equation}
For $\xi_1=\xi_2=\frac{1}{\sqrt{2}}$, the entangled state reduces to a maximally entangled state (a Bell state)
\begin{equation}
\begin{aligned}
|\Psi\rangle_{AB}=\frac{1}{\sqrt{2}}(|1\rangle_A|0\rangle_B-|0\rangle_A|1\rangle_B).
\label{eq32}
\end{aligned}
\end{equation}

The complete process can be described as follows.

$t_0$:$\star$ $A$ and $B$ are in the Bell state:
\begin{equation}
\begin{aligned}
|\Psi\rangle_{AB}&=\frac{1}{\sqrt{2}}(|1\rangle_A|0\rangle_B-|0\rangle_A|1\rangle_B)\\
&=\frac{1}{\sqrt{2}}(|n_+\rangle_A|n_-\rangle_B-|n_-\rangle_A|n_+\rangle_B).\nonumber
\end{aligned}
\end{equation}

$t_1$:$\star$ $A$ is subject to the unitary $U_{\bm{n}}(\beta)=e^{-i\beta\frac{\bm{\sigma}\cdot\bm{n}}{2}}$.

$\phantom{t_1{:}}\star$ The joint state becomes
\begin{equation}
\begin{aligned}
\frac{1}{\sqrt{2}}(e^{-i\frac{\beta}{2}\mathbf{\sigma}\bm{n}}|n_+\rangle_A|n_-\rangle_B-
e^{-i\frac{\beta}{2}\mathbf{\sigma}\bm{n}}|n_-\rangle_A|n_+\rangle_B).\nonumber
\end{aligned}
\end{equation}

$t_2$:$\star$ The form of $\bm{n}$ is now known.

$\phantom{t_2{:}}\star$ $B$ is measured in the bases \{$|n_\pm\rangle$\}.

$\phantom{t_2{:}}\star$ $A$ collapse to $e^{-i\beta\frac{\bm{\sigma}\cdot\bm{n}}{2}}|n_\mp\rangle$ with the probability\\ $p=\xi_1^2+\xi_2^2=1$.

$t_3$:$\star$ $A$ is measured in the optimal $\{|n_\pm\rangle\}$ bases.

The measurement projects the ancilla onto \{$|n_\pm\rangle$\}. It can be interpreted as the state being effectively sent back in time and flipped into the orthogonal state \{$|n_\mp\rangle$\}, to serve as the encoding  state of the probe at $t_0$ \cite{Song2024}.

\section{Entanglement-Assisted Metrology with Large spins in Non-Orthogonal Ancilla States}\label{sec3}

In Sec.~\ref{sec2}, we  considered  the scenario where the states of ancilla are a superposition of mutually orthogonal ones, forming measurement bases for $B$. Successful measurement then retroactively prepares the probe in an optimal encoding  state and  is consequently rotated by $U_A=U_{\bm{n}}(\beta)$. The protocol required the encoding  entangled state to satisfy $_B\langle\psi_i|\psi_j\rangle_B=\delta_{ij}$ (for any $|\psi_i\rangle_B$ with non-zero $c_i$).

In this section, we relax this requirement and consider the general case where the ancilla states ${|\psi_i\rangle_B}$ are not necessarily mutually orthogonal. From Eq.~(\ref{eq15}), we have known
\begin{equation}
\begin{aligned}
|\Psi\rangle_{AB}&=\sum_{i=1}^{m}|\phi_i\rangle_A\otimes|\tilde{\psi}_i\rangle_B=\sum_{i=1}^{m}c_i|\phi_i\rangle_A\otimes
|\psi_i\rangle_B,\\
|\tilde{\psi}_i\rangle_B&=\sum_{k=1}^{m}\xi_k S_{ki}|v_k\rangle_B.
\label{eq35}
\end{aligned}
\end{equation}
We can no longer simply use ${|\psi_i\rangle}_B$ as the measurement basis if the condition $_B\langle\psi_i|\psi_j\rangle_B = \delta_{ij}$ fails to hold for some or all $i$ and $j$. In this case, the entangled state after the unitary interaction can be further expressed as
\begin{equation}
\begin{aligned}
U_A|\Psi\rangle_{AB}&=\sum_{i=1}^{m-1}c_iU_A|\phi_i\rangle_A|\psi_i\rangle_B+
c_mU_A|n_-\rangle_A|\psi_m\rangle_B\\
&=c_1U_A|n_+\rangle_A|\psi_1\rangle_B+
\sum_{i=2}^{m}c_iU_A|\phi_i\rangle_A|\psi_i\rangle_B.
\label{eq36}
\end{aligned}
\end{equation}
If we  want to achieve the probe $A$ collapsing into the optimal state $U_A|n_-\rangle_A$, we have to construct a measurement basis $|\varphi\rangle$  satisfying the following condition,
\begin{equation}
\begin{aligned}
\langle\varphi|\psi_i\rangle_B=0.
\label{eq37}
\end{aligned}
\end{equation}
$|\psi_i\rangle_B$ in Eq.~(\ref{eq37}) are states with non-zero $c_i$ for $i=1\cdots m-1$. Then one measures ancilla $B$, and conditionally on the outcome of $B$ discards or keeps (postselects) the probe $A$.

The protocol proceeds as follows: Starting from the general entangled state $|\Psi\rangle_{AB}=\sum_{k=1}^{m}\xi_k|u_k\rangle_A\otimes|v_k\rangle_B$, we apply the unitary $U_{\bm{n}}(\beta)$ to the probe. The ancilla $B$ is then measured in the bases \{$|\varphi\rangle,\{|\varphi^\perp\rangle\}$\} after $\bm n$ is revealed. \{$|\varphi^\perp\rangle$\} are states which form orthogonal and complete bases together with $|\varphi\rangle$. If the outcome is $|\varphi\rangle$, the experiment proceeds with  the probe $A$. Otherwise, the run is discarded. This postselection procedure, upon success, prepares the probe in the desired optimal state $U_A|n_-\rangle_A$. The success probability for this postselection is given by
\begin{equation}
\begin{aligned}
p=c_m^2|\langle\varphi|\psi_m\rangle|^2.
\label{eq38}
\end{aligned}
\end{equation}

The measurement basis $|\varphi\rangle$ satisfying Eq.~(\ref{eq37}) may not be unique. Denoting the space satisfying  Eq.~(\ref{eq37}) by $S$, we now explore  the particular solution that maximizes the success probability for the postselection. To proceed, we decompose $|\psi_m\rangle$ into two orthogonal components
\begin{equation}
\begin{aligned}
|\psi_m\rangle=P_S|\psi_m\rangle+(I-P_S)|\psi_m\rangle.
\label{eq39}
\end{aligned}
\end{equation}
$P_S|\psi_m\rangle$ is the projection of $|\psi_m\rangle$ in the solution space $S$. $(I-P_S)|\psi_m\rangle$ is the component of $|\psi_m\rangle$ in the space orthogonal to $S$.
For any $|\varphi\rangle\in S$,
\begin{equation}
\begin{aligned}
\langle\varphi|\psi_m\rangle=\langle\varphi|P_S|\psi_m\rangle+\langle\varphi|(I-P_S)|\psi_m\rangle.
\label{eq40}
\end{aligned}
\end{equation}
Since $|\varphi\rangle$ is in $S$ and $(I-P_S)|\psi_m\rangle$ is in the space orthogonal to $S$, therefore $\langle\varphi|(I-P_S)|\psi_m\rangle=0$. Hence
\begin{equation}
\begin{aligned}
\langle\varphi|\psi_m\rangle=\langle\varphi|P_S|\psi_m\rangle.
\label{eq41}
\end{aligned}
\end{equation}
According to the Cauchy-Schwartz inequality \cite{Stee2004},
\begin{equation}
\begin{aligned}
|\langle\varphi|P_S|\psi_m\rangle|\leq\parallel\varphi\parallel\cdot\parallel P_S|\psi_m\rangle\parallel=
\parallel P_S|\psi_m\rangle\parallel.
\label{eq42}
\end{aligned}
\end{equation}
The condition for the equal sign to hold is that $|\varphi\rangle$ is parallel to $P_S|\psi_m\rangle$, that is
\begin{equation}
\begin{aligned}
|\varphi\rangle=\frac{P_S|\psi_m\rangle}{\parallel{P_S|\psi_m\rangle}\parallel}.
\label{eq43}
\end{aligned}
\end{equation}
With this consideration,  we might choose $|\varphi\rangle$ in Eq.~(\ref{eq43}) as the measurement basis for a given  $|\psi_m\rangle$, since it  maximizes the success probability $p$. If the probe $A$ collapses to the optimal encoding state  $U_A|n_+\rangle_A$, keep the run.
\begin{figure}[t]
	\centering
	\includegraphics[width=0.5\textwidth]{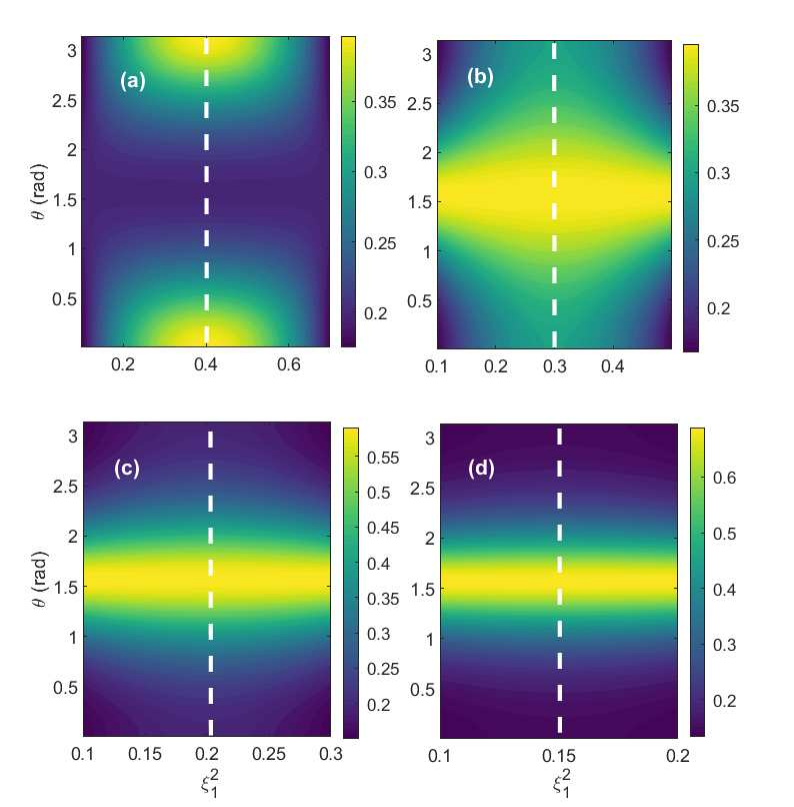}
	\caption {The probability of successfully collapsing to the metrologically optimal state $U_A|n_-\rangle_A$ for the probe $A$ as a function of $\theta$ and $\xi_1^2$ under several different $\xi_2^2$. (a) $\xi_2^2=0.2$. (b) $\xi_2^2=0.4$. (c) $\xi_2^2=0.6$. (d) $\xi_2^2=0.7$. Due to $\xi_1^2+\xi_2^2+\xi_3^2=1$, when $\xi_1$ and $\xi_2$ are fixed, $\xi_3$ will naturally be fixed as well. The white dotted lines correspond to the case that $\xi_1=\xi_3$.}
	\label{aqbpic2}
\end{figure}

To be concrete,  we take the spin-1 probe as an example. For the spin-1 probe, the Hamiltonian reads
\begin{equation}
\begin{aligned}
H=\boldsymbol{S}\cdot\boldsymbol{n}=S_x\text{sin}\theta+S_z\text{cos}\theta.
\label{eq44}
\end{aligned}
\end{equation}
The spin matrices for spin-1 take
\begin{equation}
S_x
=\frac{\hbar}{\sqrt2}
\begin{pmatrix}
0 & 1 & 0\\
1 & 0 & 1\\
0 & 1 & 0
\end{pmatrix},
S_z
=\hbar
\begin{pmatrix}
1 & 0 & 0\\
0 & 0 & 0\\
0 & 0 & -1
\end{pmatrix}.
\label{eq45}
\end{equation}
The eigenvalues of $H$ are $E_1=1, E_2=0, E_3=-1$. The corresponding eigenstates are

\begin{equation}
\small
|E_1\rangle
=
\begin{pmatrix}
\text{cos}^2\frac{\theta}{2} \\
\frac{1}{\sqrt2}\text{sin}\theta\\
\text{sin}^2\frac{\theta}{2}
\end{pmatrix},
|E_2\rangle
=
\begin{pmatrix}
-\frac{1}{\sqrt2}\text{sin}\theta \\
\text{cos}\theta\\
\frac{1}{\sqrt2}\text{sin}\theta
\end{pmatrix},
|E_3\rangle
=
\begin{pmatrix}
\text{sin}^2\frac{\theta}{2} \\
-\frac{1}{\sqrt2}\text{sin}\theta\\
\text{cos}^2\frac{\theta}{2}
\end{pmatrix}.
\label{eq46}
\end{equation}
The optimal probe states $|n_\pm\rangle$ can be given  using Eq.~(\ref{eq46}), and the transformation matrix in the form of Eq.~(\ref{eq14}) reads
\begin{equation}
\begin{pmatrix}
|0\rangle \\
|1\rangle \\
|2\rangle
\end{pmatrix}
=
\begin{pmatrix}
\frac{1}{\sqrt{2}} & -\frac{\text{sin}\theta}{\sqrt{2}} & \frac{\text{cos}\theta}{\sqrt{2}}\\
0 & \text{cos}\theta & \text{sin}\theta\\
\frac{1}{\sqrt{2}} & \frac{\text{sin}\theta}{\sqrt{2}} & -\frac{\text{cos}\theta}{\sqrt{2}}
\end{pmatrix}
\begin{pmatrix}
|n_+\rangle \\
|E_2\rangle\\
|n_-\rangle
\label{eq47}
\end{pmatrix}.
\end{equation}

From Eq.~(\ref{eq12}), we choose the bases as $\{|u_1\rangle=|0\rangle, |u_2\rangle=|1\rangle, |u_3\rangle=|2\rangle\}$ for the probe, and $\{|v_1\rangle=|0\rangle, |u_2\rangle=|1\rangle, |u_3\rangle=|2\rangle\}$ for the ancilla. At time $t_0$, the encoding  general entangled state is given by
\begin{equation}
\begin{aligned}
|\Psi\rangle_{AB}=\xi_1|0\rangle_A|0\rangle_B+\xi_2|1\rangle_A|1\rangle_B+\xi_3|2\rangle_A|2\rangle_B.
\label{eq48}
\end{aligned}
\end{equation}
The Schmidt coefficients satisfy the normalization condition $\xi_1^2+\xi_2^2+\xi_3^2=1$. For the entangled state, at least two of $\xi_1$, $\xi_2$ and $\xi_3$ must be not zero. After the unitary $U_A$ on $A$, the joint state becomes
\begin{equation}
\begin{aligned}
U_A|\Psi\rangle_{AB}&=c_1U_A|n_+\rangle_A|\psi_1\rangle_B+c_2U_A|E_2\rangle_A|\psi_2\rangle_B\\&+
c_3U_A|n_-\rangle_A|\psi_3\rangle_B,
\label{eq49}
\end{aligned}
\end{equation}
where
\begin{equation}
\begin{aligned}
|\tilde{\psi}_1\rangle_{B}&=\frac{\xi_1}{\sqrt{2}}|0\rangle_B+\frac{\xi_3}{\sqrt{2}}|2\rangle_B,\\
|\tilde{\psi}_2\rangle_{B}&=-\frac{\xi_1\text{sin}\theta}{\sqrt{2}}|0\rangle_B+\xi_2\text{cos}\theta|1\rangle_B+
\frac{\xi_3\text{sin}\theta}{\sqrt{2}}|2\rangle_B,\\
|\tilde{\psi}_3\rangle_{B}&=\frac{\xi_1\text{cos}\theta}{\sqrt{2}}|0\rangle_B+\xi_2\text{sin}\theta|1\rangle_B-
\frac{\xi_3\text{cos}\theta}{\sqrt{2}}|2\rangle_B.
\label{eq50}
\end{aligned}
\end{equation}
Here we consider the case where all $c_i>0$. This requires $\xi_2\neq0$. It is easy to find  that   $_B\langle\psi_i|\psi_j\rangle_B = \delta_{ij}$ does not always hold for any  $i$ and $j$. To ensure that probe $A$ is ultimately projected into the optimal state $U_A|n_-\rangle_A$, we need to construct the measurement basis vector $|\varphi\rangle$  satisfying  $\langle\varphi|\psi_{1(2)}\rangle=0$. After simple algebra, we find the form of $|\varphi\rangle$
\begin{equation}
\begin{aligned}
|\varphi\rangle&=\alpha_0|0\rangle+\alpha_1|1\rangle+\alpha_2|2\rangle,\\
\alpha_0&=\frac{\xi_2\xi_3\text{cos}\theta}{\sqrt{2}N},\alpha_1=\frac{\xi_1\xi_3\text{sin}\theta}{N},
\alpha_2=-\frac{\xi_1\xi_2\text{cos}\theta}{\sqrt{2}N},
\label{eq51}
\end{aligned}
\end{equation}
where $N=[\frac{1}{2}\xi_2^2(\xi_1^2+\xi_3^2)\text{cos}^2\theta+\xi_1^2\xi_3^2\text{sin}^2\theta]^{1/2}$. The special case where $N=0$ will be analyzed in the Appendix \ref{appendixA}. After the unitary interaction, $\bm n$ is known, and one can measures $B$ in the basis \{$|\varphi\rangle,\{|\varphi^\perp\rangle\}$\}.  If the outcome is $|\varphi\rangle$, the experiment proceeds with a measurement on the probe. From Eq.~(\ref{eq38}), the probability of postselecting successfully is
\begin{equation}
\begin{aligned}
p=\frac{\xi_2^2}{\frac{\xi_2^2(1-\xi_2^2)}{2\xi_1^2\xi_3^2}\text{cos}^2\theta+\text{sin}^2\theta}.
\label{eq52}
\end{aligned}
\end{equation}

The postselected state of the probe $A$ is $U_A|n_-\rangle_A$. Fig.~\ref{aqbpic2} plots the success probability $p$ as a function of the polar angle $\theta$ and the coefficient $\xi_1^2$, for different fixed values of $\xi_2^2$. For small $\xi_2^2$ [Fig.~\ref{aqbpic2}(a)], $p$ is relatively larger near $\theta=0$ and $\theta=\pi$, but smaller in between. As $\xi_2^2$ increases [Figs.~\ref{aqbpic2}(b)-(d)], the maximum of $p$ shifts toward $\theta=\pi/2$, and the peak becomes narrower. The white dotted line in Fig.~\ref{aqbpic2} (corresponding to $\xi_1=\xi_3$) yields a higher success probability $p$ than any other case for all $\theta$. This maximum can be explained from Eq.~(\ref{eq52}): for fixed $\xi_2^2$ and $\theta$, we have $\xi_1^2\xi_3^2\leq(\frac{\xi_1^2+\xi_3^2}{2})^2=(\frac{1-\xi_2^2}{2})^2$. Equality holds  and then $p$ is maximized, when $\xi_1=\xi_3=\sqrt{\frac{1-\xi_2^2}{2}}$. Thus, we can get the maximum of the successful probability for different $\theta$
\begin{equation}
\begin{aligned}
p_\text{max}=\frac{\xi_2^2(1-\xi_2^2)}{2\xi_2^2\text{cos}^2\theta+\text{sin}^2\theta(1-\xi_2^2)}.
\label{eq54}
\end{aligned}
\end{equation}
When $\xi_1=\xi_3$, $\langle\psi_1|\psi_{2(3)}\rangle=0$. This orthogonality allows us to use \{$|\varphi\rangle, |\psi_1\rangle, |\psi_2\rangle$\} as the measurement bases for $B$ after $\bm n$ is known. A measurement on $B$ then yields two outcomes of interest: If the result is $|\varphi\rangle$, probe $A$ collapses to $U_A|n_-\rangle_A$, and if it is $|\psi_1\rangle$, $A$ collapses to $U_A|n_+\rangle_A$. The total success probability $P$ in this case is
\begin{equation}
\begin{aligned}
P=\frac{\xi_2^2(1-\xi_2^2)}{2\xi_2^2\text{cos}^2\theta+\text{sin}^2\theta(1-\xi_2^2)}+\frac{1-\xi_2^2}{2}.
\label{eq55}
\end{aligned}
\end{equation}
\begin{figure}[t]
	\centering
	\includegraphics[width=0.48\textwidth]{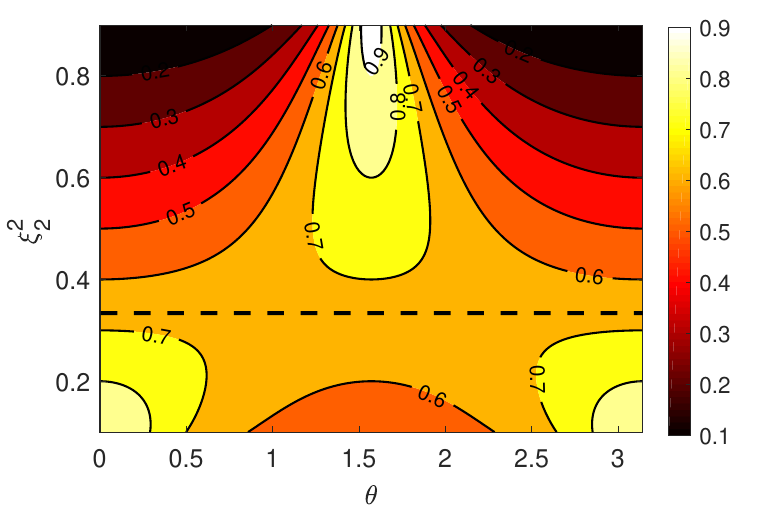}
	\caption {Contour plot of total successful probability $P$ over the parameter space $\xi_2^2-\theta$. $\xi_1=\xi_3=\sqrt{(1-\xi_2^2)/2}$. The dotted line  corresponds to the maximally entangled case $\xi_1=\xi_2=\xi_3=\frac{1}{\sqrt{3}}$, for which $|\psi_1\rangle=|n_+\rangle$, $|\psi_2\rangle=|E_2\rangle$, $|\psi_3\rangle=|n_-\rangle$ and satisfies $_B\langle\psi_i|\psi_j\rangle_B = \delta_{ij}$.}
	\label{aqbpic3}
\end{figure}

The dependence of the total success probability on $\xi_2^2$ and $\theta$ is plotted in Fig.~\ref{aqbpic3}. It is observed  that even for a general (non-maximally) entangled encoding  state, there remains a finite success probability to prepare an optimal probe for any $U_{\bm{n}}(\beta)$ ($\bm n$-agnostic) by making a reasonable choice of the encoding  state. The probability depends on $\bm n$ and the specific form of the entangled state.  This special case recovers the protocol discussed in Sec.~\ref{sec2a} for a maximally entangled encoding  state of spin-1 particles with success probability $p=\frac{2}{3}$.

Discussions on the postselection is now in order.
Postselection  can significantly enhance the performance of classical parameter estimation. The main idea is to filter the  measurement outputs before the subsequent data processings. By doing so, it not only improves the Fisher information obtainable from a single measurement event but also boosts the information efficiency of the final data processing procedures. For quantum parameter estimation, these enhancements are largely constrained in general, because the protocols yielding higher precision are rarely obtained due to a lower probability of successful postselection. It is shown that this precision can further be improved with the help of quantum resources like entanglement and negativity in the quasiprobability distribution\cite{Arvi2020}. However, these quantum advantages in attaining considerable success probability with large precision are bounded irrespective of any accessible quantum resources\cite{das2023}. Although the postselection is not always helpful in quantum parameter estimation, yet this work is interesting as it contributes to the agnostic parameter estimation theory.

\section{CONCLUSIONS}\label{sec4}
In this manuscript, we propose a proposal to employ  the advantage of large spin  for enhancing the agnostic parameter estimation precision. Our results shows that, even without prior knowledge of the generator of the unitary operator, maximum quantum Fisher information can be achieved by utilizing entanglement between the large spin probe and the ancilla. Two cases are considered. First, we consider the case that the state of the  ancilla is a superposition of  mutually orthogonal states. A crucial difference from the previous studies of qubit is that achieving optimal estimation with large spins requires postselection, which succeeds with a certain probability in most cases not one. By applying this scheme, we analyze the case where the encoding  state is a maximally entangled state between the probe and ancilla spins. We have obtained the explicit dependence of the postselection success probability on the spin quantum number $m$. Second, beyond the  maximally entangled states, we demonstrate that optimal estimation remains achievable even when the probe and ancilla are prepared in general (non-maximal) entangled state. However, the success probability in this  case depends on both the specific form of the entangled state and the generator $H$.
We believe that with the advancement of current quantum technology, the optical set ups \cite{Arvi2023} can provide a feasible experimental platform to realize  the present proposal.

\section*{ACKNOWLEDGMENTS}\label{sec5}
This work is supported by National Natural Science Foundation of China (NSFC) under Grant No. 12575010.

\appendix

\section{Some special cases in Sec.~\ref{sec3}}\label{appendixA}
\begin{figure}[b]
	\centering
	\includegraphics[width=0.43\textwidth]{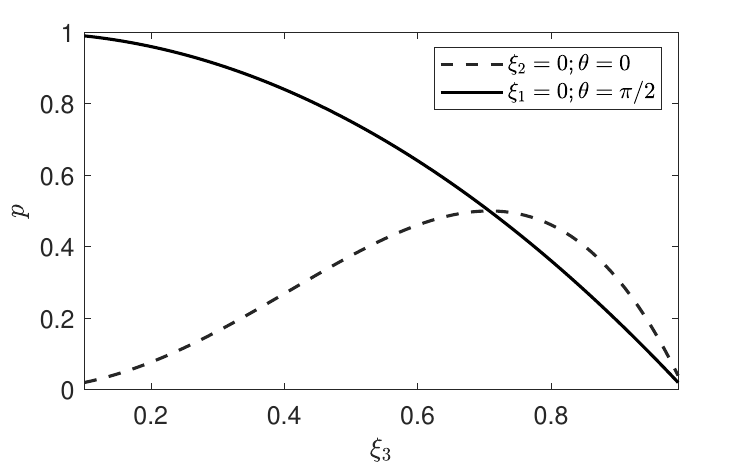}
	\caption {The probability of successfully collapsing to the metrologically optimal state $U_A|n_-\rangle_A$ for the probe $A$ as a function of $\xi_3$. The dotted line corresponds to the case that $\xi_2=0, \theta=0$. The solid line corresponds to the case that $\xi_1=0, \theta=\pi/2$.}
	\label{aqbpic4}
\end{figure}
Eq.~(\ref{eq51}) shows that $N=0$ yields $\xi_2=0$ and $\theta=0$ or $\theta=\pi$; $\xi_1=0$ or $\xi_3=0$ and $\theta=\pi/2$. When $\xi_2=0$, $|\tilde{\psi}_2\rangle_B=-\frac{\sin\theta}{\cos\theta}|\tilde{\psi}_3\rangle_B$. The probe cannot collapse to $U_A|n_-\rangle_A$ unless $\theta=0$ or $\theta=\pi$.  We first consider the case that $\xi_2=0$ and $\theta=0$ or $\theta=\pi$. From Eq.~(\ref{eq50}), we get
\begin{equation}
\begin{aligned}
|\psi_1\rangle_{B}&=\xi_1|0\rangle_B+\xi_3|2\rangle_B,\\
|\psi_3\rangle_{B}&=\xi_1|0\rangle_B-\xi_3|2\rangle_B.
\label{eqa1}
\end{aligned}
\end{equation}
We need to construct the measurement basis vector $|\varphi\rangle$ such that $\langle\varphi|\psi_{1}\rangle=0$. Combining this condition with Eq.~(\ref{eq43}), we get the form of $|\varphi\rangle$
\begin{equation}
\begin{aligned}
|\varphi\rangle&=-\xi_3|0\rangle+\xi_1|2\rangle.
\label{eqa2}
\end{aligned}
\end{equation}
The probability of successfully collapsing to the metrologically optimal state $U_A|n_-\rangle_A$ is
\begin{equation}
\begin{aligned}
p=|c_3|^2|\langle\varphi|\psi_3\rangle|^2=2\xi_3^2(1-\xi_3^2).
\label{eqa3}
\end{aligned}
\end{equation}
The dependence of the success probability on $\xi_3$ is plotted in Fig.~\ref{aqbpic4}. It is clear that when $\xi_1=\xi_3$, $p$ reaches its maximum value $p=1/2$. Meanwhile, when $\xi_1=\xi_3$, $_B\langle\psi_3|\psi_1\rangle_B=0$. This allows us to use $|\psi_1\rangle_B$ and $|\psi_3\rangle_B$ as the measurement bases for $B$ once $\bm n$ is known. If the measurement outcome is $|\psi_1\rangle_B$, probe $A$ collapses to $U_A|n_+\rangle_A$; if it is $|\psi_3\rangle_B$, $A$ collapses to $U_A|n_-\rangle_A$.  In this case, the overall success probability becomes $p=1$. This special case corresponds to the protocol discussed in Sec.~\ref{sec2b}, where the adjustable coefficients [Eq.~(\ref{eq27})] in the encoding  entangled state coincidentally match those required for the unknown $\bm n$. In this case the success probability reaches its maximum of 1. Similarly, when $\theta=\pi$, the same result will be obtained.

Another special case is $\theta=\frac{\pi}{2}$, $\xi_1=0$ or $\xi_3=0$. From Eq.~(\ref{eq50}), when $\xi_1=0$, we get $|\psi_1\rangle_{B}=|\psi_2\rangle_{B}=|2\rangle_B$ and $|\psi_3\rangle_{B}=|1\rangle_B$. Because $\langle\psi_3|\psi_{1(2)}\rangle=0$, this orthogonality allows us to use \{$|0\rangle_B, |1\rangle_B, |2\rangle_B$\} as the measurement bases for $B$ after $\bm n$ is known. If the results is $|1\rangle$, the probe collapses to $U_A|n_-\rangle_A$. The success probability in this case is
\begin{equation}
\begin{aligned}
p=\xi_2^2=1-\xi_3^2.
\label{eqa4}
\end{aligned}
\end{equation}
From Fig.~\ref{aqbpic4}, the probability approaches 1 as $\xi_2$ tends to 1. This particular case arises when $\theta=\pi/2$, the optimal encoding  state $|n_-\rangle_A$ just equal to $|1\rangle$. Consequently, at $\xi_2=1$, the probe is initialized entirely in this optimal state. Similarly, when $\xi_3=0$, the same result would be obtained.


\begin{thebibliography}{0}
\expandafter\ifx\csname natexlab\endcsname\relax\def\natexlab#1{#1}\fi
\expandafter\ifx\csname bibnamefont\endcsname\relax
  \def\bibnamefont#1{#1}\fi
\expandafter\ifx\csname bibfnamefont\endcsname\relax
  \def\bibfnamefont#1{#1}\fi
\expandafter\ifx\csname citenamefont\endcsname\relax
  \def\citenamefont#1{#1}\fi
\expandafter\ifx\csname url\endcsname\relax
  \def\url#1{\texttt{#1}}\fi
\expandafter\ifx\csname urlprefix\endcsname\relax\def\urlprefix{URL }\fi
\providecommand{\bibinfo}[2]{#2}
\providecommand{\eprint}[2][]{\url{#2}}

\end{thebibliography}


\begin{thebibliography}{10}

\bibitem{Gio2006}V. Giovannetti, S. Lloyd, and L. Maccone, Quantum Metrology, Phys. Rev. Lett. \textbf{96}, 010401 (2006).

\bibitem{Gio2011}V. Giovannetti, S. Lloyd, and L. Maccone, Advances in quantum metrology, Nat. Photon. \textbf{5}, 222 (2011).

\bibitem{Degen2017}C. L. Degen, F. Reinhard, and P. Cappellaro, Quantum sensing, Rev. Mod. Phys. \textbf{89}, 035002 (2017).

\bibitem{Liu2020}J. Liu, H. Yuan, X.-M. Lu, and X. Wang, Quantum fisher information matrix and multiparameter estimation, J. Phys. A: Math. Theor. \textbf{53}, 023001 (2020).

\bibitem{Ali2014}S. Alipour, M. Mehboudi, and A. T. Rezakhani, Quantum Metrology in Open Systems: Dissipative Cram$\acute{\text{e}}$r-Rao Bound, Phys. Rev. Lett. \textbf{112}, 120405 (2014).

\bibitem{Hou2020}Z. Hou, Z. Zhang, G.-Y. Xiang, C.-F. Li, G.-C. Guo, H. Chen, L. Liu, and H. Yuan, Minimal Tradeoff and Ultimate Precision Limit of Multiparameter Quantum Magnetometry under the Parallel Scheme, Phys. Rev. Lett. \textbf{125}, 020501 (2020).

\bibitem{Hum2013}P. C. Humphreys, M. Barbieri, A. Datta, and I. A. Walmsley, Quantum Enhanced Multiple Phase Estimation, Phys. Rev. Lett. \textbf{111}, 070403 (2013).

\bibitem{Dorn2009}U. Dorner, R. Demkowicz-Dobrza$\acute{\text{n}}$ski, B. J. Smith, J. S. Lundeen, W. Wasilewski, K. Banaszek, and I. A. Walmsley, Optimal Quantum Phase Estimation, Phys. Rev. Lett. \textbf{102}, 040403 (2009).

\bibitem{Yuan2015}H. Yuan and C.-H. F. Fung, Optimal Feedback Scheme and Universal Time Scaling for Hamiltonian Parameter Estimation, Phys. Rev. Lett. \textbf{115}, 110401 (2015).

\bibitem{Chen2024}H. Chen, Y. Chen, J. Liu, Z. Miao, H. Yuan, Quantum Metrology Enhanced by Leveraging Informative Noise with Error Correction, Phys. Rev. Lett. \textbf{133}, 190801 (2024).

\bibitem{Lu2021}X.-M. Lu and X. Wang, Incorporating Heisenberg's Uncertainty Principle into Quantum Multiparameter Estimation, Phys. Rev. Lett. \textbf{126}, 120503 (2021).

\bibitem{Ding2021}W. Ding, X. Wang, and S. Chen, Fundamental Sensitivity Limits for Non-Hermitian Quantum Sensors, Phys. Rev. Lett. \textbf{131}, 160801 (2023).

\bibitem{Jing2015}X.-X. Jing, J. Liu, H.-N. Xiong, and X. Wang, Maximal quantum Fisher information for general su(2) parametrization processes, Phys. Rev. A \textbf{92}, 012312 (2015).

\bibitem{Bai2023}S.-Y. Bai and J.-H. An, Floquet Engineering to Overcome No-Go Theorem of Noisy Quantum Metrology, Phys. Rev. Lett. \textbf{131}, 050801 (2023).

\bibitem{Peng2024}J.-X. Peng, B. Zhu, W. Zhang, and K. Zhang, Dissipative quantum Fisher information for a general Liouvillian parametrized process, Phys. Rev. A \textbf{109}, 012432 (2024).

\bibitem{Yu2023}X. Yu and C. Zhang, Quantum parameter estimation of non-Hermitian systems with optimal measurements, Phys. Rev. A \textbf{108}, 022215 (2023).

\bibitem{Dem2014}R. Demkowicz-Dobrza$\acute{\text{n}}$ski and L. Maccone, Using Entanglement Against Noise In Quantum Metrology, Phys. Rev. Lett. \textbf{113}, 250801 (2014).

\bibitem{Li2024}J. Li, D. Cui, and X. X. Yi, Parameter estimation with limited access to measurements, Phys. Rev. A \textbf{109}, 032215 (2024).

\bibitem{Li2023}J. Li, H. Liu, Z. Wang, and X. X. Yi, Enhanced parameter estimation by measurement of non-Hermitian operators, AAPPS Bull. \textbf{33}, 22 (2023).

\bibitem{Shao2023}L. Shao, R. Zhang, W. Lu, Z. Zhang, and X. Wang, Quantum phase transition in the XXZ central spin model, Phys. Rev. A \textbf{107}, 013714 (2023).

\bibitem{Pang2014}S. Pang and T. A. Brun, Quantum metrology for a general Hamiltonian parameter, Phys. Rev. A \textbf{90}, 022117 (2014).

\bibitem{Beau2017}M. Beau and A. del Campo, Nonlinear Quantum Metrology of Many-Body Open Systems, Phys. Rev. Lett. \textbf{119}, 010403 (2017).

\bibitem{Arvi2023}D. R. M. Arvidsson-Shukur, A. G. McConnell, and N. Yunger Halpern, Nonclassical Advantage in Metrology Established via Quantum Simulations of Hypothetical Closed Timelike Curves, Phys. Rev. Lett. \textbf{131}, 150202 (2023).

\bibitem{Song2024}X. Song, F. Salvati, C. Gaikwad, N. Yunger Halpern, D. R. M. Arvidsson-Shukur, and K. Murch, Agnostic Phase Estimation, Phys. Rev. Lett. \textbf{132}, 260801 (2024).

\bibitem{Song2025}X. Song, S. S. Borjigin, F. Salvati, Y.-X. Wang, N. Y. Halpern, D. R. M. Arvidsson-Shukur, and K. Murch, Superconducting antiqubits achieve optimal phase estimation via unitary inversion, arXiv:2506.04315.

\bibitem{Hel1976}C. W. Helstrom, \textit{Quantum Detection and Estimation Theory} (Academic Press, New York, 1976).

\bibitem{Hol1982}A. S. Holevo, \textit{Probabilistic and Statistical Aspects of Quantum Theory} (North-Holland, Amsterdam, 1982).

\bibitem{Hub1992}M. H$\ddot{\text u}$bner, Explicit computation of the Bures distance for density matrices, Phys. Lett. A \textbf{163}, 239 (1992)

\bibitem{Hub1993}M. H$\ddot{\text u}$bner, Computation of Uhlmann's parallel transport for density matrices and the Bures metric on three-dimensional Hilbert space, Phys. Lett. A \textbf{179}, 226 (1993).

\bibitem{Brau1994}S. L. Braunstein and C. M. Caves, Statistical distance and the geometry of quantum states, Phys. Rev. Lett. \textbf{72}, 3439 (1994).

\bibitem{Brau1996}S. L. Braunstein, C. M. Caves, and G. J. Milburn, Generalized uncertainty relations: Theory, examples, and lorentz invariance, Ann. Phys. \textbf{247}, 135 (1996).

\bibitem{Cramer1946}H. Cram$\acute{\text{e}}$r, \textit{Mathematical Methods of Statistics} (Princeton University, Princeton, NJ, 1946), p. 500.

\bibitem{Godel1949}K. G$\ddot{\text o}$del, An example of a new type of cosmological solutions of Einstein's field equations of gravitation, Rev. Mod. Phys. \textbf{21}, 447 (1949).

\bibitem{Mor1988}M. S. Morris, K. S. Thorne, and U. Yurtsever, Wormholes, time machines, and the weak energy condition, Phys. Rev. Lett. \textbf{61}, 1446 (1988).

\bibitem{Svet2011}G. Svetlichny, Time travel: Deutsch vs. teleportation, Int. J. Theor. Phys. \textbf{50}, 3903 (2011).

\bibitem{Lloyd2011}S. Lloyd, L. Maccone \textit{et al}., Closed timelike curves via postselection: Theory and experimental test of consistency, Phys. Rev. Lett. \textbf{106}, 040403 (2011).

\bibitem{Lloy2011}S. Lloyd, L. Maccone, R. Garcia-Patron, V. Giovannetti, and Y. Shikano, Quantum mechanics of time travel through post-selected teleportation, Phys. Rev. D \textbf{84}, 025007 (2011).

\bibitem{Aha2002}Y. Aharonov and L. Vaidman, \textit{Time in Quantum Mechanics} (Springer, New York, 2002), pp. 369-412.

\bibitem{Sch1907}E. Schmidt, Math. Ann. \textbf{63}, 433 (1907); A. Ekert and P. L. Knight, Am. J. Phys. \textbf{63}, 415 (1995).

\bibitem{Pere1995}A. Peres, \textit{Quantum Theory: Concepts and Methods} (Kluwer Academic Publishers, Dordrecht, 1995); A. Peres, Phys. Lett. A \textbf{202}, 16 (1995).

\bibitem{Stee2004}J. M. Steele, \textit{The Cauchy-Schwarz Master Class: An Introduction to the Art of Mathematical Inequalities} (Cambridge University Press, Cambridge, England, 2004).

\bibitem{Arvi2020}D. R. M. Arvidsson-Shukur, N. Y. Halpern, H. V. Lepage, A. A. Lasek, C. H. W. Barnes, and S. Lloyd, Quantum advantage in postselected metrology, Nature communications \textbf{11}, 3775 (2020), see also arXiv: 1903.02563 (2019).

\bibitem{das2023}  S. Das, S. Modak, M.  N. Bera, Bounding Quantum Advantages in Postselected Metrology, Phys. Rev. A \textbf{107}, 042413 (2023).



\end{thebibliography}
\end{document}